\documentclass[english,graybox]{svmult}
\usepackage[colorlinks=true,urlcolor=black,linkcolor=black,citecolor=black]
{hyperref}

\usepackage[T1]{fontenc}
\usepackage[latin9]{inputenc}
\usepackage{amsmath}
\usepackage{amssymb}
\usepackage{amsfonts}
\usepackage{babel}
\usepackage{graphicx}%

\usepackage{mathptmx}       
\usepackage{helvet}         
\usepackage{courier}        
\usepackage{type1cm}        
\usepackage{makeidx}         
\usepackage{graphicx}        
\usepackage{multicol}        
\usepackage[bottom]{footmisc}

\makeindex             

\begin{document}

\title*{Asymptotically flat black holes and gravitational waves
 in three-dimensional massive gravity}
\titlerunning{Asymptotically flat black holes and gravitational waves in 3D massive gravity}  
\author{ Cédric Troessaert, David Tempo and Ricardo Troncoso}
\institute{ 
Cédric Troessaert \at Centro de Estudios Científicos (CECs), Av. Arturo Prat 514, Valdivia, Chile, \email{troessaert@cecs.cl} \and David Tempo , \at  Universit{\'e} Libre de Bruxelles and International Solvay Institutes ULB-Campus Plaine C.P.231, B-1050 Brussels, Belgium,   \at  Centro de Estudios Científicos (CECs), Av. Arturo Prat 514, Valdivia, Chile \email{tempo@cecs.cl} \and
Ricardo Troncoso \at Centro de Estudios Científicos (CECs), Av. Arturo Prat 514, Valdivia, Chile, \email{troncoso@cecs.cl}}
%
%

\maketitle

\abstract{Different classes of exact solutions for the BHT massive gravity
theory are constructed and analyzed. We focus in the special case
of the purely quadratic Lagrangian, whose field equations are irreducibly
of fourth order and are known to admit a unique (flat) maximally symmetric
solution, as well as asymptotically locally flat black holes endowed
with gravitational hair. The first class corresponds to a Kerr-Schild
deformation of Minkowski spacetime along a covariantly constant null
vector. As in the case of General Relativity, the field equations
linearize so that the generic solution can be easily shown to be described
by four arbitrary functions of a single null coordinate. These solutions
can then be regarded as a new sort of pp-waves. The second class is
obtained from a deformation of the static asymptotically locally flat
black hole, that goes along the spacelike (angular) Killing vector.
Remarkably, although the deformation is not of Kerr-Schild type, the
field equations also linearize, and hence the generic solution can
be readily integrated. It is neither static nor spherically symmetric,
being described by two integration constants and two arbitrary functions
of the angular coordinate. In the static case it describes ``black
flowers\textquotedblright{} whose event horizons break the spherical
symmetry. The generic time-dependent solution appears to describe
a graviton that moves away from a black flower. Despite the asymptotic
behaviour of these solutions at null infinity is relaxed with respect
to the one for General Relativity, the asymptotic symmetries coincide.
However, the algebra of the conserved charges corresponds to BMS$_{3}$,
but devoid of central extensions. The ``dynamical black flowers\textquotedblright{}
are then shown to possess a finite energy. The surface integrals that
define the global charges also turn out to be useful in the description
of the thermodynamics of solutions with event horizons.}

\section{Introduction}

After a century, General Relativity still remains as one of the banners
of what theoretical physicists regard as a heuristic theory. It has
also raised one the biggest challenges of the discipline, being the
suitable description of the gravitational field at a microscopic level.
Although the problem is unsolved yet, there have been different interesting
proposals, including supergravity, string theory, loop quantum gravity
and the gauge/gravity correspondence (for reviews about these subjects,
see e.g., \cite{Reviews:Supergravity-1,Reviews:Supergravity-2,StringT-1,StringT-2,StringT-3,StringT-4,StringT-5,StringT-6,LQG-1,LQG-2,LQG-3,AdS/CFT-1,AdS/CFT-2,AdS/CFT-3}).
Some of them actually pursue the more ambitious goal of constructing
a single unified theory that described matter and their interactions.
Since none of the proposals are still fully satisfactory, some part
of the community has in parallel tried to follow a more modest approach,
focusing on toy models that capture some of the key features of the
gravitational field. Along this line, the same theory of General Relativity,
once formulated on three spacetime dimensions turns out to be much
simpler than its four-dimensional counterpart, and consequently, many
things have been learned. For instance, in the case of negative cosmological
constant $\Lambda$, black holes in vacuum were shown to be obtained
just from suitable identifications of the maximally symmetric AdS$_{3}$
background \cite{BTZ}, \cite{BHTZ}. This simplicity then appears
to be inherited in Strominger's proposal for the microscopic counting
of the asymptotic growth of the number of states of a black hole \cite{Strominger-Cardy},
which allows to recover the semiclassical Bekenstein-Hawking black
hole entropy in terms of Cardy formula \cite{Cardy}. The idea relies
on an old result by Brown and Henneaux \cite{BrownHenneaux}, which
states that the asymptotic symmetries of General Relativity with $\Lambda<0$
in three spacetime dimensions are spanned by the algebra of the two-dimensional
conformal group with a precise central extension. This naturally suggests
that a quantum theory of gravity in 3D would be described in terms
of a conformal field theory in two dimensions. Nowadays, this result
clearly appears as a precursor of the AdS/CFT correspondence \cite{MaldacenaLargeN,GubserKlebanovPolyakovGT,WittenAdSHolo}.
Within this set up, there are interesting proposals to quantize the
theory, as the one in \cite{WittenRevisited,MaloneyWittenQG3D}, which
relies on assuming that the partition function can be holomorphically
factorized. Recent different approaches have also been presented in
e.g., \cite{MaloneyMicroStates3D,KimPorratiCanonicalQG3D}. Despite
the circle has not yet been completed, the lessons learned from three-dimensional
General Relativity appear to provide new and good hints towards the
solution of the problem in four spacetime dimensions. However, for
the sake of the main topic of this book, that concerns gravitational
waves, this toy model is certainly oversimplified because of the absence
of local degrees of freedom. In other words, General Relativity in
three dimensions has no chance of possessing propagating gravitons,
and so from this point of view, the theory turns out to be trivial.
This is a direct consequence of a purely geometrical fact: in three
dimensions the Weyl tensor identically vanishes, and hence the Riemann
tensor can be expressed in terms of the Ricci tensor and the curvature
scalar. Thus, according to the Einstein equations, the curvature becomes
completely determined by the stress-energy tensor, so that in vacuum,
spacetime is necessarily of constant curvature $\Lambda$.

It is then interesting to look for different models of gravity in
three-dimensional spacetimes that could incorporate local propagating
degrees of freedom for the gravitational field. In order to do that,
it is useful to recall some remarks about Riemannian geometry in three
dimensions. Indeed, unlike the case of higher-dimensional spaces,
the fact that the Weyl tensor identically vanishes does not mean that
three-dimensional metrics are conformally flat. Indeed, in 3D, the
suitable object that plays this role is the Cotton tensor 
\begin{equation}
C_{\mu\nu}=\frac{1}{\sqrt{g}}\varepsilon_{\mu}^{\ \alpha\beta}\nabla_{\alpha}\left(R_{\beta\nu}-\frac{1}{4}g_{\beta\nu}R\right)\ ,\label{Cotton tensor}
\end{equation}
being symmetric and traceless. Thus, for any three-dimensional conformally
flat spacetime the Cotton tensor vanishes, while the converse is also
true but just on local patches. This tensor is a fundamental piece
of the ``topologically massive gravity\textquotedblright \ theory
of Deser, Jackiw and Templeton \cite{Deser-Jackiw-Templeton-1,Deser-Jackiw-Templeton-2},
whose Lagrangian is the one of General Relativity plus an additional
term proportional to the three-dimensional Lorentz-Chern-Simons form.
The field equations in vacuum are given by 
\begin{equation}
G_{\mu\nu}+\Lambda g_{\mu\nu}-\frac{1}{\mu}C_{\mu\nu}=0\ ,\label{TMG-field-eqs}
\end{equation}
which by virtue of (\ref{Cotton tensor}) are of third order for the
metric and not invariant under parity. Nonetheless, and remarkably,
upon linearization around a maximally symmetric background solution,
it is simple to verify that the theory exorcises the ghosts that generically
appear in theories with higher order field equations, since it possesses
a healthy single propagating degree of freedom. Note that this sort
of massive graviton differs from the Fierz-Pauli one, which possesses
two degrees of freedom and respects parity. It is worth mentioning
that since constant curvature spacetimes are conformally flat, they
trivially solve the field equations of topologically massive gravity
(\ref{TMG-field-eqs}), and therefore, in the case of $\Lambda<0$,
BTZ black holes do. The theory in vacuum also admits interesting exact
solutions for which spacetime is not of constant curvature, see e.g.,
\cite{warped-AdS-1,warped-AdS-2,warped-AdS-3,warped-AdS-C,pp-waves,warped-AdS-AWPS,kundt,warped-AdS with gravitons,Laura-Gaston,Gaston-Yerko}.

More recently, a different theory of massive gravity in three spacetime
dimensions has been proposed by Bergshoeff, Hohm and Townsend (BHT)
\cite{BHT-1,BHT-2}. The action is described by the Einstein-Hilbert
one with cosmological constant, plus the addition of a precise combination
of quadratic terms in the curvature, as in (\ref{Action-purely quadratic}),
but with an independent coupling. The field equations are then of
fourth order and invariant under parity. Also noteworthy, once the
field equations are linearized on a maximally symmetric background,
they reduce to the ones of Fierz and Pauli for a massive graviton.

In the full nonlinear case, the field equations in vacuum are known
to admit a wide class of interesting exact solutions, see e.g., \cite{ExactSolutionsBHT-1,ExactSolutionsBHT-4,BHT-2,OTT-BHs-1,ExactSolutionsBHT-6,ExactSolutionsBHT-7,ExactSolutionsBHT-5,ExactSolutionsBHT-8,ExactSolutionsBHT-10,ExactSolutionsBHT-11,ExactSolutionsBHT-12,ExactSolutionsBHT-SUPER}.

It is worth pointing out that, as it occurs for the Einstein-Gauss-Bonnet
theory in $d>4$ dimensions, which is also quadratic in the curvature
\cite{Boulware-Deser,BH-Scan}, the theory actually admits two different
maximally symmetric solutions in vacuum, unless the couplings are
chosen in a special way, so that the theory acquires an interesting
behaviour.

In the special case in which the BHT theory admits a unique maximally
symmetric vacuum of constant curvature $\lambda$, at the perturbative
level there is a single local propagating degree of freedom, being
described by a ``partially massless graviton\textquotedblright{}
\cite{PartiallyMassless-2,PartiallyMassless-1,PartiallyMassless-3,PartiallyMassless-4,Deser-Alas,BHT-2}.
This special case, in the limit of vanishing $\lambda$, is known
to enjoy very remarkable properties \cite{Deser-Alas}, and it is
then the main focus of the remaining sections.

\section{Asymptotically locally flat black holes in BHT massive gravity}

Hereafter, we consider the BHT massive gravity theory in the case
that admits a unique maximally symmetric (flat) background solution.
The action is then described just by the terms that are purely quadratic
in the curvature, which reads 
\begin{equation}
I\left[g\right]=\frac{1}{16\pi G}\int d^{3}x\sqrt{-g}\left(R_{\mu\nu}R^{\mu\nu}-\frac{3}{8}R^{2}\right)\ ,\label{Action-purely quadratic}
\end{equation}
so that the field equations become irreducibly of fourth order, and
are given by 
\begin{equation}
2\square R_{\mu\nu}-\frac{1}{2}\nabla_{\mu}\nabla_{\nu}R-\frac{1}{2}\square Rg_{\mu\nu}+4R_{\mu\sigma\nu\rho}R^{\sigma\rho}-\frac{3}{2}RR_{\mu\nu}-R_{\sigma\rho}R^{\sigma\rho}g_{\mu\nu}+\frac{3}{8}R^{2}g_{\mu\nu}=0\;.\label{FEqs}
\end{equation}

This three-dimensional gravity theory possesses an additional peculiar
feature: unlike the case of General Relativity or topologically massive
gravity, it admits asymptotically locally flat black holes in vacuum
\cite{OTT-BHs-1}. In the static case, the metric is given by 
\begin{equation}
ds^{2}=-\left(br-\mu\right)dt^{2}+\frac{dr^{2}}{br-\mu}+r^{2}d\phi^{2}\ ,\label{BH-static}
\end{equation}
so that the Ricci scalar reads 
\begin{equation}
R=-\frac{2b}{r}\ .
\end{equation}
The geometry then possesses a spacelike singularity at the origin,
which in the case of $b>0$ and $\mu>0$, is cloaked by the event
horizon located at $r_{+}=\mu/b$. Note that if $\mu=0$, there is
a NUT right on top of the singularity which becomes null. This class
of black holes turns out to be asymptotically locally flat because,
in spite of the fact that the metric does not approach the Minkowski
one in the asymptotic region, the Riemann tensor vanishes anyway as
$r\rightarrow\infty$. It is also worth pointing out that spacetimes
of the form (\ref{BH-static}) are conformally flat. Indeed, these
black hole solutions were first found in the context of conformal
gravity in vacuum, for which the field equations imply that the Cotton
tensor (\ref{Cotton tensor}) vanishes \cite{Joao-Pessoa}\footnote{In this sense, it is amusing to see that the asymptotically locally
flat black hole (\ref{BH-static}) can be constructed from the conformal
gluing of three copies of BTZ\ black holes at infinity, so that the
asymptotically locally flat region is mapped to the horizon of two
of them. Besides, the conformal flatness of this class of black holes
makes them also to fulfill the field equations of the Poincaré gauge
theory \cite{Blagojevic}.}.

Note that the Hawking temperature 
\begin{equation}
T=\frac{b}{4\pi}\ ,\label{Temperature-static}
\end{equation}
does not depend on the integration constant $\mu$ in (\ref{BH-static}).
As we show below, the integration constant $b$ turns out to be related
to the black hole mass, which is the only nonvanishing conserved charge.
Hence, the remaining integration constant $\mu$ can be interpreted
as a gravitational hair parameter. It should be emphasized that dealing
with the nonstandard fall off of a metric as in (\ref{BH-static}),
together with the fact that the field equations are of fourth order
and quadratic in the curvature, makes the construction of the surface
integrals that describe the corresponding conserved charges to be
a somehow cumbersome task. This is explicitly carried out for a generic
set up in sections \ref{Asymptotic-Conditions-Symmetries} and \ref{Surface-Charges}.

In order to catch a glimpse of the answer, one can readily compute
the black hole entropy through Wald's approach \cite{Wald-Entropy},
which turns out to be given by 
\begin{equation}
S=\frac{\pi b}{4G}\ .\label{Entropy-static}
\end{equation}
differing from ``a quarter of the area\textquotedblright \ of the
event horizon. Nevertheless, this is not surprising because what one
expects for General Relativity clearly does not have to apply for
the theory under discussion. The mass can then be found if one assumes
that the first law of thermodynamics holds, which must be the case
for any suitable action principle, i.e., 
\begin{equation}
\delta M=T\delta S\ ,
\end{equation}
and hence the mass can be integrated as 
\begin{equation}
M=\frac{b^{2}}{32G}\ .\label{Mass-static}
\end{equation}

\bigskip{}

The following sections are devoted to the discussion of different
classes of exact solutions, including an interesting new type of pp-waves,
as well as ``dynamical black flowers\textquotedblright{} \cite{BT3-Arxiv}.
The asymptotic behaviour of the theory that accommodates solutions
of the latter class is also analyzed, including the explicit construction
of the surface integrals that define the corresponding conserved charges
and their algebra. Some aspects of the thermodynamics of the solutions
with event horizons are also explored.

\section{Partially massless pp-waves on flat spacetime}

Let us consider flat Minkowski spacetime in null coordinates 
\begin{equation}
d\bar{s}^{2}=2dudv+dx^{2}\ ,\label{Flat-Minkowski-metric}
\end{equation}
as the seed background solution of the field equations (\ref{FEqs})
to be generalized. Following Kerr and Schild \cite{Kerr-Schild-1,Kerr-Schild-2},
one might look for a consistent deformation of (\ref{Flat-Minkowski-metric})
along a null Killing vector, that for simplicity is also assumed to
be covariantly constant. The spacetime metric can then be written
as 
\begin{equation}
ds^{2}=d\bar{s}^{2}+H(u,x)du^{2}\ ,\label{pp-wave metric}
\end{equation}
where the function $H$ parametrizes the deformation. Thus, despite
the field equations (\ref{FEqs}) are quadratic in the curvature,
as in the case of four-dimensional General Relativity, they also exactly
linearize without making any sort of approximation. Indeed, they just
reduce to 
\begin{equation}
\partial_{x}^{4}H=0\ ,\label{pp-wave eq}
\end{equation}
which readily integrates so that the function that describes the deformation
acquires a generic form given by 
\begin{equation}
H(u,x)=A_{0}(u)+A_{1}(u)x+A_{2}(u)x^{2}+A_{3}(u)x^{3}\ ,\label{H-pp-wave}
\end{equation}
depending on four arbitrary functions $A_{I}(u)$. Therefore, the
theory turns out to admit exact solutions for which the spacetime
metric is given by (\ref{pp-wave metric}) with (\ref{H-pp-wave}),
that appear to describe a wide class of pp-waves.

It is interesting to check that the Kerr-Schild approach turns out
to be successful in order to generate exact solutions beyond General
Relativity. Results along these lines have been previously obtained
for topologically massive gravity \cite{kundt,warped-AdS with gravitons,Laura-Gaston,Gaston-Yerko},
as well as for the generic BHT massive gravity theory \cite{ExactSolutionsBHT-6,KC-Anzats-BHT-1}.
This is also the case for different theories in higher dimensions
whose field equations are quadratic in the curvature, as the one described
by the Einstein-Gauss-Bonnet action in vacuum \cite{KC-Anzats-EGB-1,KC-Anzats-EGB-2,KC-Anzats-Lovelock-3,KC-Anzats-Lovelock-4}.

One may naturally wonder whether new solutions could be obtained by
regarding the black hole (\ref{BH-static}) instead of Minkowski spacetime
(\ref{Flat-Minkowski-metric}) as the seed background solution. This
is the main subject of the next section.

\section{Dynamical black flowers}

\label{DynamicalBlackFlowers}

Let us now regard the asymptotically locally flat black hole (\ref{BH-static})
as the background seed solution. It turns out to be useful to write
the seed black hole metric in terms of a set of coordinates that is
suitably adapted to null infinity. Defining the tortoise coordinate
$r^{\ast}$ according to $dr^{\ast}=\frac{dr}{br-\mu}$, one then
sets $u=t-r^{\ast}$, so that the black hole solution in (\ref{BH-static})
acquires the form 
\begin{equation}
ds^{2}=-\left(br-\mu\right)du^{2}-2dudr+r^{2}d\phi^{2}\ .\label{StaticMetric}
\end{equation}
We thus look for a different class of deformations as compared with
the one in the previous section. Indeed, the deformation is assumed
to go along the spacelike Killing vector $\partial_{\phi}$, so that
the metric reads 
\begin{equation}
ds^{2}=-\left(br-\mu\right)du^{2}-2dudr+\left(r-\mathcal{H}\left(u,\phi\right)\right)^{2}d\phi^{2}\;,\label{eq:AnzatDBF}
\end{equation}
where $\mathcal{H}\left(u,\phi\right)$ stands for an arbitrary function
that is periodic in $\phi$. Quite remarkably, although the deformation
is not of Kerr-Schild type, the field equations (\ref{FEqs}) also
exactly linearize without the need of any approximation. Indeed, once
(\ref{eq:AnzatDBF}) is plugged within (\ref{FEqs}), the field equations
just reduce to a single one that is linear in $\mathcal{H}$, which
reads 
\begin{equation}
\partial_{u}\left(\partial_{u}\mathcal{H}+\frac{b}{2}\mathcal{H}\right)=0\;.\label{eq:LDE}
\end{equation}
The general solution can then be readily integrated, so that the function
that parametrizes the deformation reads 
\begin{equation}
\mathcal{H}\left(u,\phi\right)=\mathcal{A}\left(\phi\right)+\mathcal{B}\left(\phi\right)e^{-\frac{b}{2}u}\;,\label{eq:H-DBF}
\end{equation}
where $\mathcal{A}\left(\phi\right)$ and $\mathcal{B}\left(\phi\right)$
stand for arbitrary periodic functions of the angular coordinate.
It is worth pointing out that once the deformation is switched on,
the solution of the form (\ref{eq:AnzatDBF}) with (\ref{eq:H-DBF})
remains being conformally flat. Besides, its Ricci scalar reads 
\begin{equation}
R=\frac{2b}{r-\mathcal{H}}\;,
\end{equation}
which manifestly signals the existence of a curvature singularity
located at $r=\mathcal{H}(u,\phi)$, and hence one can set the range
of the radial coordinate according to $\mathcal{H}<r<\infty$.

This class of solutions is then naturally divided in two cases: \textbf{(i)}
$\mathcal{B}\left(\phi\right)=0$, so that spacetime is static, and
\textbf{(ii)} $\mathcal{B}\left(\phi\right)\neq0$, leading to metrics
that describe dynamical spacetimes.\\

In case (i) the static metric given by (\ref{eq:AnzatDBF}) and (\ref{eq:H-DBF}),
with $\mathcal{B}\left(\phi\right)=0$, possesses an event horizon
at $r=r_{+}=\mu/b$. Cosmic censorship then requires the horizon to
surround the curvature singularity at the origin of the radial coordinate,
and hence the function $\mathcal{A}\left(\phi\right)$ turns out to
be bounded from above according to: 
\[
\mathcal{A}\left(\phi\right)<r_{+}\ .
\]
\ The shape of the deformed horizon is then generically no longer
spherically symmetric. Indeed, the induced metric on $r=r_{+}$ and
a constant value of the coordinate $u$ is given by $g_{\phi\phi}=\left(r_{+}-\mathcal{A}\left(\phi\right)\right)^{2}$,
so that for an arbitrary periodic arbitrary function $\mathcal{A}\left(\phi\right)$,
spacetime appears to describe a sort of ``black flower\textquotedblright .
Note that despite the horizon has been deformed, rigidity still holds,
and hence the Hawking temperature of the black flowers agrees with
the one of the undeformed spherically symmetric case, given by (\ref{Temperature-static}).

In the dynamical case (ii), since the additional arbitrary periodic
function does not vanish, i.e., for $\mathcal{B}\left(\phi\right)\neq0$,
the deformation $\mathcal{H}\left(u,\phi\right)$ propagates along
outgoing null rays. Hence, the generic solution is neither static
nor spherically symmetric, and it is clear that at late retarded time
$u\rightarrow\infty$, the configuration tends to a static black flower.
Therefore, this class of solutions can be regarded as a black flower
endowed with an outgoing graviton. The behaviour of this class of
``dynamical black flowers\textquotedblright{} somehow resembles the
one of Robinson-Trautman spacetimes for General Relativity in four
dimensions \cite{RobinsonTrautman-1,RobinsonTrautman-2,About-RobinsonTrautman-1,About-RobinsonTrautman-2},
which describe outgoing radiation on a Schwarzschild black hole. Nonetheless,
a difference that should be stressed is that the evolution in time
for these four-dimensional analogues is described by an involved fourth-order
nonlinear differential equation. Further details about the behaviour
of this class of dynamical black flowers can be found in \cite{BT3-Arxiv}.

As a closing remark of this section, we would like to mention that
the strategy of finding exact solutions for deformations that are
non null, and then not of Kerr-Schild type (see e. g. \cite{KS-Extended-2}),
has also been successful in the context of a generic BHT massive gravity
theory \cite{KC-Anzats-BHT-1} as well as in the case of Lovelock
theories in higher dimensions \cite{KS-Extended-1}.

\section{Relaxed asymptotically flat behaviour and asymptotic symmetries}

\label{Asymptotic-Conditions-Symmetries}

Here we explore the asymptotically locally flat behaviour of the theory
described by the action (\ref{Action-purely quadratic}) that is able
to accommodate the class of ``dynamical black flowers\textquotedblright \ discussed
in the previous section. As it has been the main focus of attention
for a big deal of the recent literature, see, e.g.,
 \cite{BMSaspect-1,Barnich:2006av,Barnich:2012aw,Bagchi:2012yk,CosmoEntro-1,FlatHolography-1,Bagchi:2013lma,Barnich:2013axa,stromingerBMS-2,stromingerBMS-1,StromingerBMS-3,Strominger:2014pwa,Barnich:2015mui}, our description is made so that the region where the asymptotic
symmetries and their corresponding global charges are defined, is
located towards null infinity. Following a generic criterion (see,
e.g., \cite{HenneauxTeitelboim-AdSspaces,HMTZ}), we assume that the
fall off of the gravitational field should be mapped into itself at
least under the Poincaré group. Furthermore, the asymptotic behaviour
must be relaxed enough such that the solutions of interest, in this
case given by the dynamical black flowers, fit within the set; but
at the same time, the fall-off has to be sufficiently fast so as to
yield a finite set of global conserved charges.

The deviations with respect to the flat Minkowski reference background,
$\Delta g_{\mu\nu}=g_{\mu\nu}-\bar{g}_{\mu\nu}$, with 
\begin{equation}
d\bar{s}^{2}=-du^{2}-2dudr+r^{2}d\phi^{2}\ ,
\end{equation}
are then proposed to be asymptotically of the form 
\begin{align}
\Delta g_{uu} & =h_{uu}r+f_{uu}+\text{ }\cdot\cdot\cdot\nonumber \\
\Delta g_{u\phi} & =h_{u\phi}r+f_{u\phi}+\text{ }\cdot\cdot\cdot\nonumber \\
\Delta g_{\phi\phi} & =h_{\phi\phi}r+f_{\phi\phi}+\text{ }\cdot\cdot\cdot\label{BoundaryConditions}\\
\Delta g_{rr} & =\frac{k_{rr}}{r^{3}}+\text{ }\cdot\cdot\cdot\nonumber \\
\Delta g_{ur} & =\frac{k_{ur}}{r}+\text{ }\cdot\cdot\cdot\nonumber \\
\Delta g_{\phi r} & =\frac{k_{r\phi}}{r}+\text{ }\cdot\cdot\cdot\nonumber 
\end{align}
where $h_{\mu\nu},$ $f_{\mu\nu}$ and $k_{\mu\nu}$ stand for arbitrary
functions of $u$ and $\phi$. As it can be seen from (\ref{BoundaryConditions}),
the presence of terms along the functions $h_{\mu\nu}$, makes the
fall off of the metric to be clearly relaxed as compared with the
one for General Relativity in three-dimensional spacetimes at null
infinity \cite{ABS-AS-Null-3D-GR,GC-Null-Conditions-3D}. Nonetheless,
the asymptotic symmetries, being spanned by diffeomorphisms that maintain
the asymptotic form of the metric, i. e., that fulfill 
\[
\mathcal{L}_{\xi}g_{\mu\nu}=\mathcal{O}(\Delta g_{\mu\nu})\ ,
\]
remain the same. Indeed, the asymptotic behaviour of the metric in
(\ref{BoundaryConditions}) is mapped into itself under diffeomorphisms
spanned by asymptotic Killing vectors whose leading terms are given
by 
\begin{align}
\xi^{u} & =T\left(\phi\right)+u\partial_{\phi}Y\left(\phi\right)+\text{ }\cdot\cdot\cdot\nonumber \\
\xi^{r} & =-r\partial_{\phi}Y\left(\phi\right)+\text{ }\cdot\cdot\cdot\label{Asympt-KVs}\\
\xi^{\phi} & =Y\left(\phi\right)-\frac{1}{r}\partial_{\phi}\left(T\left(\phi\right)+u\partial_{\phi}Y\left(\phi\right)\right)+\text{ }\cdot\cdot\cdot\ .\nonumber 
\end{align}
Diffeomorphisms of this sort, for which $T(\phi)=Y(\phi)=0$ then
form an ideal of ``pure gauge\textquotedblright{} transformations,
so that the quotient algebra defines the so-called BMS$_{3}$ one,
given by the semidirect sum of a Virasoro algebra with an abelian
ideal that corresponds to supertranslations.

It is worth to highlight that relaxing the asymptotic behaviour as
in (\ref{BoundaryConditions}) does not spoil the asymptotic symmetries
that are known to appear in the simpler case of pure General Relativity.

\section{Conserved charges from surface integrals at null infinity and black
hole thermodynamics}

\label{Surface-Charges}

One of the main aims of this section is the explicit construction
of conserved charges that correspond to the asymptotic symmetries
spanned by (\ref{Asympt-KVs}). This task is fairly simplified once
the action (\ref{Action-purely quadratic}) is expressed in terms
of an auxiliary field $\mathcal{\ell}_{\mu\nu}$, which allows to
reduce the field equations from fourth to second order \cite{BHT-2,Deser-Alas}.
The action (\ref{Action-purely quadratic}) can then be written as
\begin{equation}
I\left[g,\mathcal{\ell}\right]=\frac{1}{16\pi G}\int d^{3}x\sqrt{-g}\left(\mathcal{\ell}^{\mu\nu}G_{\mu\nu}-\frac{1}{4}\left(\mathcal{\ell}^{\mu\nu}\mathcal{\ell}_{\mu\nu}-\mathcal{\ell}^{2}\right)\right)\ ,\label{Igl}
\end{equation}
so that field equations associated to $\mathcal{\ell}_{\mu\nu}$ turn
out to be algebraic and read 
\begin{equation}
G^{\mu\nu}-\frac{1}{2}\left(\mathcal{\ell}^{\mu\nu}-g^{\mu\nu}\mathcal{\ell}\right)=0\ .
\end{equation}
Consequently, the auxiliary field $\mathcal{\ell}_{\mu\nu}$ is given
by 
\begin{equation}
\mathcal{\ell}_{\mu\nu}=2\left(R_{\mu\nu}-\frac{1}{4}g_{\mu\nu}R\right)\ ,\label{Lmunu}
\end{equation}
being proportional to the Schouten tensor\footnote{Note that the Schouten tensor can be regarded as a sort of ``potential\textquotedblright{}
for the Cotton tensor in (\ref{Cotton tensor}).}. The variation of the action (\ref{Igl}) with respect to the metric
then yields second order field equations 
\begin{multline}
E_{\mu\nu}=\nabla^{\alpha}\nabla_{\alpha}\mathcal{\ell}^{\mu\nu}-2\nabla_{\rho}\nabla^{(\mu}\mathcal{\ell}^{\nu)\rho}+\nabla^{\mu}\nabla^{\nu}\mathcal{\ell}+g^{\mu\nu}\left(\nabla_{\rho}\nabla_{\lambda}\mathcal{\ell}^{\rho\lambda}-\nabla^{\alpha}\nabla_{\alpha}\mathcal{\ell}\right)+4\mathcal{\ell}^{\lambda(\mu}G_{\hspace{0.05in}\lambda}^{\nu)}\\
+\mathcal{\ell}^{\mu\nu}R-\mathcal{\ell}R^{\mu\nu}-g^{\mu\nu}\mathcal{\ell}^{\rho\sigma}G_{\rho\sigma}+\mathcal{\ell}^{\lambda\mu}\mathcal{\ell}_{\hspace{0.05in}\lambda}^{\nu}-\mathcal{\ell\ell}^{\mu\nu}-\frac{1}{4}g^{\mu\nu}\left(\mathcal{\ell}_{\alpha\beta}\mathcal{\ell}^{\alpha\beta}-\mathcal{\ell}^{2}\right)=0\;.\label{FEqs-2nd-order}
\end{multline}

One can then follow the covariant approach in \cite{Covariant-Barnich-Brandt,GC-Boundary-Charges},
so that the conserved charges are the form 
\begin{equation}
Q_{\xi}=\int_{0}^{1}ds\left(\frac{1}{2}\int_{\partial\Sigma}\varepsilon_{\nu\mu\alpha}\tilde{k}_{\xi}^{\left[\nu\mu\right]}dx^{\alpha}\right)\ ,\label{Charges}
\end{equation}
where $\tilde{k}_{\xi}^{\left[\nu\mu\right]}$ stands for the ``superpotential\textquotedblright ,
whose explicit from is somewhat involved and can be found in \cite{BT3-Arxiv}.
For the sake of this lecture, it is enough to mention that once the
asymptotic form of the gravitational field in (\ref{BoundaryConditions})
is taken into account, it is possible to use an arbitrary set of interpolating
metrics that fulfill the asymptotic conditions, so that the conserved
charges (\ref{Charges}) associated to the asymptotic symmetries spanned
by (\ref{Asympt-KVs}) reduce to 
\begin{equation}
Q_{\xi}=Q\left[T,Y\right]=\frac{1}{64\pi G}\int d\phi\left(\left(T+u\partial_{\phi}Y\right)h_{uu}^{2}+2Yh_{u\phi}h_{uu}+4\partial_{\phi}Yh_{uu}+4Y\partial_{u}h_{u\phi}\right)\ .\label{Charges-Asympt}
\end{equation}

The algebra of the conserved charges (\ref{Charges-Asympt}) can then
be directly read from the variation of the same surface integrals,
because $\left\{ Q_{\xi_{1}},Q_{\xi_{2}}\right\} :=\delta_{\xi_{2}}Q_{\xi_{1}}$.
Thus, considering that the leading nontrivial terms of the asymptotic
field equations (\ref{FEqs-2nd-order}) are given by 
\begin{align}
E_{uu} & =-\frac{1}{4}\left(\partial_{u}h_{uu}^{2}\right)r^{-1}+\mathcal{O}\left(r^{-2}\right)=0\ ,\\
E_{u\phi} & =\frac{1}{4}\left(\partial_{\phi}h_{uu}^{2}-2\partial_{u}\left(h_{uu}h_{u\phi}\right)+4\partial_{u}\left(\partial_{\phi}h_{uu}-\partial_{u}h_{u\phi}\right)\right)r^{-1}+\mathcal{O}\left(r^{-2}\right)=0\ ,
\end{align}
together with the transformation law of the relevant components of
the asymptotic form of the metric under the asymptotic symmetries
(\ref{Asympt-KVs}), that read 
\begin{align}
\delta_{\xi}h_{uu} & =\left(T\left(\phi\right)+u\partial_{\phi}Y\left(\phi\right)\right)\partial_{u}h_{uu}+\partial_{\phi}\left(h_{uu}Y\right)\;,\\
\delta_{\xi}h_{u\phi} & =h_{uu}\partial_{\phi}\left(T\left(\phi\right)+u\partial_{\phi}Y\left(\phi\right)\right)+\left(T\left(\phi\right)+u\partial_{\phi}Y\left(\phi\right)\right)\partial_{u}h_{u\phi}+\partial_{\phi}\left(h_{u\phi}Y\right)\;,
\end{align}
it is found that that the algebra of the conserved charges precisely
agrees with the one of the asymptotic symmetries. Indeed, the algebra
does not admit central extensions, since 
\begin{equation}
\left\{ Q_{\xi_{1}},Q_{\xi_{2}}\right\} -Q_{\left[\xi_{1},\xi_{2}\right]}\approx\mathcal{K}_{_{\xi_{1},\xi_{2}}}\approx0\ .
\end{equation}
Once expanded in Fourier modes 
\begin{equation}
\mathcal{P}_{n}=Q\left[e^{in\phi},0\right]\;;\;\;\mathcal{J}_{n}=Q\left[0,e^{in\phi}\right]\;,
\end{equation}
the algebra acquires the standard form, given by 
\begin{align}
i\left\{ \mathcal{J}_{m},\mathcal{J}_{n}\right\}  & =\left(m-n\right)\mathcal{J}_{m+n}\;,\nonumber \\
i\left\{ \mathcal{J}_{m},\mathcal{P}_{n}\right\}  & =\left(m-n\right)\mathcal{P}_{m+n}\;,\label{BMS3-Algebra}\\
i\left\{ \mathcal{P}_{m},\mathcal{P}_{n}\right\}  & =0\;.\nonumber 
\end{align}
The absence of central extensions agrees with the claim in \cite{FareghbalHosseini},
which relies on the vanishing cosmological constant limit of the centrally
extended two-dimensional conformal algebra that was found in \cite{PTT}
(see also \cite{OTT-BHs-1} ) for BHT massive gravity in the special
case (with a unique maximally symmetric vacuum) from a holographic
approach.

\subsection{Global charges of (rotating) black holes and dynamical black flowers}

Since the global charges associated to spacetimes whose asymptotic
behaviour is described by (\ref{BoundaryConditions}) have already
been found to be given by (\ref{Charges-Asympt}), one can readily
evaluate them for the solutions of interest.

\bigskip{}

\textit{Static asymptotically locally flat black holes.-} The mass
of the solution described by the line element in (\ref{BH-static})
is directly obtained from (\ref{Charges-Asympt}), by taking into
account that the only relevant nonvanishing deviation from the flat
background is given by $h_{uu}=-b$. The suitable surface integral
that reproduces the expected result for the mass is then given by,
\begin{equation}
M=Q\left(\partial_{u}\right)=\frac{b^{2}}{32G}\ ,\label{MQ1}
\end{equation}
in agreement with the result obtained in eq. (\ref{Mass-static})
of the introduction, that came from integrating the first law of thermodynamics.
It is worth pointing out that (\ref{MQ1}) is actually the only nonvanishing
global charge (\ref{Charges-Asympt}) associated to the full set of
asymptotic symmetries in (\ref{Asympt-KVs}). Therefore, the remaining
integration constant $\mu$ can indeed be interpreted as a gravitational
hair parameter.

\bigskip{}

\textit{Asymptotically flat rotating black holes.- }The theory described
by the action (\ref{Action-purely quadratic}) has also been shown
to admit asymptotically locally flat rotating black hole solutions
\cite{OTT-BHs-1,GOTT-BHs-2}. It is useful to express the solution
in the set of coordinates of ref. \cite{PTT}, so that once written
in the outgoing null coordinates defined in section \ref{DynamicalBlackFlowers},
the metric reads 
\begin{equation}
ds^{2}=-NFdu^{2}-2N^{\frac{1}{2}}dudr+\left(r^{2}+r_{0}^{2}\right)\left(d\phi+N^{\phi}du\right)^{2}\ ,\label{RotatingBH}
\end{equation}
with 
\begin{equation}
F=br-\mu\text{\ };\text{\ }N=\frac{\left(8r+a^{2}b\right)^{2}}{64\left(r^{2}+r_{0}^{2}\right)}\text{\ };\text{\ }N^{\phi}=-\frac{a}{2}\left(\frac{br-\mu}{r^{2}+r_{0}^{2}}\right)\text{\ };\text{\ }r_{0}^{2}=\frac{a^{2}}{4}\left(\mu+\frac{a^{2}b^{2}}{16}\right)\text{ }.
\end{equation}
Note that when the rotation parameter $a$ vanishes, the spacetime
metric reduces to the static black hole solution in (\ref{BH-static}).
In this case, the relevant deviations with respect to the flat background
are given by $h_{uu}=-b$ and $h_{u\phi}=-\frac{a}{2}b$, and hence
the mass and the angular momentum can be directly obtained by virtue
of (\ref{Charges-Asympt}). They are explicitly given by 
\begin{align}
M=Q\left(\partial_{u}\right) & =\frac{b^{2}}{32G}\ ,\label{M-rot}\\
J=Q\left(\partial_{\phi}\right) & =\left(\frac{b^{2}}{32G}\right)a=Ma\ ,\label{J-rot}
\end{align}
which is in full agreement with what one recovers from the vanishing
cosmological constant limit of BHT massive gravity in the special
case that admits a unique maximally symmetric vacuum \cite{PTT,FareghbalHosseini}.
It is worth pointing out that the integration constant $\mu$ remains
as a gravitational hair parameter, because it is simple to verify
that the mass and the angular momentum in (\ref{M-rot}) and (\ref{J-rot})
are indeed the only nonvanishing global charges associated to the
full set of asymptotic symmetries.

\bigskip{}

\textit{Dynamical black flowers.- }In this case, despite spacetime
is neither spherically symmetric nor static, the solution described
by (\ref{eq:AnzatDBF}), (\ref{eq:H-DBF}) also fits within the set
of asymptotic conditions (\ref{BoundaryConditions}). It is amusing
to see that actually their global charges have already been obtained.
This is because the deformation with respect to the static black hole
does not alter the values of $h_{uu}$ and $h_{u\phi}$. Therefore,
the global charges coincide with the ones of the asymptotically locally
flat black hole, being given by $M=\frac{b^{2}}{32G}$ and $J=0$.
It should be pointed out that, even though there is outgoing radiation,
the total energy at null infinity remains constant, i.e., no ``news\textquotedblright{}
are present \cite{Bondi-News-1,Sachs-News-2}. It is also worth noting
that, once restricted to the static black flower ($\mathcal{B}\left(\phi\right)=0$),
both the constant $\mu$ and the periodic function $\mathcal{A}\left(\phi\right)$
become pure hair. In this sense, from the mode expansion, one might
interpret the static black flower as a black hole solution endowed
with an infinite number of gravitational hair parameters.

\subsection{Thermodynamics}

The generic expression for the conserved charges in (\ref{Charges})
also turns out to be useful in a different sense when event horizons
are present. Indeed, for black holes that possess a global Killing
vector that generates the horizon, like it is the case of the static
black flowers in (\ref{eq:AnzatDBF}), with (\ref{eq:H-DBF}) and
$\mathcal{B}\left(\phi\right)=0$, one can evaluate the surface charge
for $\xi=\partial_{u}$ in order to deduce some of their thermodynamical
properties. Indeed, since the superpotential $\tilde{k}_{\xi}^{[\mu\nu]}$
associated to a Killing vector $\xi$ is conserved \cite{Covariant-Barnich-Brandt,GC-Boundary-Charges},
one obtains that 
\begin{equation}
\delta\left(\left.Q_{\xi}\right\vert _{r=\infty}+\left.Q_{\xi}\right\vert _{r=r_{+}}\right)=0\ .\label{cons}
\end{equation}
For static black flowers it is remarkable that the superpotential
can be precisely evaluated at an arbitrary value of the radial coordinate.
Indeed, its only nonvanishing contribution is given by 
\begin{equation}
\tilde{k}_{\xi}^{\left[ur\right]}=\frac{b\delta b}{32\pi G}\;,
\end{equation}
which does depend neither on the radial coordinate nor on the arbitrary
function $\mathcal{A}\left(\phi\right)$. Thus, as explained above,
evaluating the global charge at infinity, one obtains the mass, i.
e., 
\begin{equation}
\left.Q_{\xi}\right\vert _{r=\infty}=\frac{b^{2}}{32G}=M\ ,
\end{equation}
while once the variation of the charge is evaluated at the horizon,
$r=r_{+}$, one finds that 
\begin{equation}
\left.\delta Q_{\xi}\right\vert _{r=r_{+}}=\int_{\Sigma_{h}}\tilde{k}_{\xi}^{\left[ur\right]}d\phi=-\frac{b\delta b}{16G}=-T\delta S\ .
\end{equation}
Therefore, the conservation law (\ref{cons}) just reduces to the
first law of thermodynamics in the canonical ensemble, which reads
$\delta\mathcal{F}=0$, where $\mathcal{F}=M-TS$ stands for the Helmholtz
free energy. One then concludes that the black hole, or generically,
the black flower entropy that comes from Wald's formula \cite{Wald-Entropy},
given by (\ref{Entropy-static}), can also be written as 
\begin{equation}
S=-2\pi\int_{0}^{1}ds\left(\int_{\Sigma_{h}}\hat{\varepsilon}_{\mu\nu}\tilde{k}_{\xi}^{\left[\nu\mu\right]}d\phi\right)\;,
\end{equation}
where $\hat{\varepsilon}_{\mu\nu}$ denotes the binormal to the bifurcation
surface $\Sigma_{h}$, and is normalized according to $\hat{\varepsilon}_{\mu\nu}\hat{\varepsilon}^{\mu\nu}=-2$.

\bigskip{}
It is also simple to verify that the results of this section naturally
extend to the case of asymptotically locally flat rotating hairy black
holes. Indeed, as a curiosity, it is worth pointing out that since
the horizon of the solution in (\ref{RotatingBH}) possesses a vanishing
angular velocity, i. e., $\Omega_{+}=0$, according to (\ref{cons}),
the first law is recovered, but in the canonical ensemble: 
\begin{equation}
dM=TdS-\Omega_{+}dJ=TdS\ .
\end{equation}

\section{Acknowledgments}

The results presented here rely on our preprint \cite{Pucon}, as
well as the more recent work in \cite{BT3-Arxiv}. We thank Oscar
Fuentealba, Hernán González, Javier Matulich and Alfredo Pérez for
enlightening discussions and especially Glenn Barnich for his valuable
collaboration. C.T. is a Laurent Houart postdoctoral fellow. R.T.
wishes to thank the Physique théorique et mathématique group of the
Université Libre de Bruxelles, and the International Solvay Institutes
for the warm hospitality. This work is partially funded by the Fondecyt
grants N${^{\circ}}$ 1130658, 1121031, 11130260, 3140125. The work
of G.B. is partially supported by research grants of the F.R.S.-FNRS
and IISN-Belgium as well as the ``Communauté française de Belgique
- Actions de Recherche Concertees\textquotedblright . The work of
D.T. is partially supported by the ERC Advanced Grant ``SyDuGraM\textquotedblright ,
by a Marina Solvay fellowship, by FNRS-Belgium (convention FRFC PDR
T.1025.14 and convention IISN 4.4514.08) and by the ``Communauté
Française de Belgique\textquotedblright \ through the ARC program.
The Centro de Estudios Científicos (CECs) is funded by the Chilean
Government through the Centers of Excellence Base Financing Program
of Conicyt.

\end{document}